\documentclass[journal]{IEEEtran}
\usepackage{blindtext}
\usepackage{graphicx}
\graphicspath{ {graph/} }
\usepackage{bm}
\usepackage{amsmath}
\usepackage{amssymb}
\usepackage{mathtools}
\usepackage[boxed]{algorithm2e}

\usepackage{cite}

\ifCLASSINFOpdf
\else
\fi
\begin{document}
%
\title{GPU acceleration of a patient-specific airway image segmentation and its assessment}
%
%
%

\author{\IEEEauthorblockN{Yu-Wei Chang$^{1}$},
\IEEEauthorblockN{Tony W.H. Sheu$^{1,2}$}\\
\IEEEauthorblockA{$^{1}$Department of Engineering Science and Ocean Engineering, National Taiwan University\\
$^{2}$Center for Advance Studies in Theoretical Science (CASTS), National Taiwan University}}

\maketitle

\begin{abstract}

\par Image segmentation plays an important role in computer vision, object detection, traffic control, and video surveillance. Typically, it is a critical step in the 3D reconstruction of a specific organ in medical image processing which unveils the detailed tomography of organ, tumor, and nerve, and thus helping to improve the quality of surgical pathology. However, there may be high computational requirements in it. With the advent of GPUs, more complex and realistic models can be simulated, but the deployment of these facilities also requires a huge amount of capital. As a consequence, how to make good use of these computational resource is essential to GPU computing.

\par This study discusses the image segmentation of 3D airway reconstruction, identifies the computing-intensive task, and parallelizes the algorithm of image segmentation in order to obtain theoretical maximum speedup in terms of the benchmark ratio of GPU to CPU. There are five steps involved, which are the image acquisition, pre-processing, segmentation, reconstruction, and object recognition. It is worth to note that it takes 85\% of time on segmentation. This study successfully accelerates the image segmentation of 3D airway reconstruction by optimizing the memory usage, grid and block setting and multiple GPUs communication, thereby gaining a total speedup of 61.8 on two GPUs (Nvidia K40).
\end{abstract}


\section{Introduction}

\par In the US, two out of every five people will be diagnosed with cancer in their lifetime. Lung cancer is the main cause of cancer mortality in man and woman\cite{RN2}. The five-year survival rate of lung cancer (17.7 percent) is dramatically lower than the other cancers, such as colon cancer (64.4 percent), breast cancer (89.7 percent), prostate cancer (98.9 percent)\cite{RN3}. Despite of the high mortality rate of lung cancer, researchers have found that early detection of such a fatal cancer by means of screening, such as low-dose CT scans\cite{RN112}, can actually increase the five-year survival rate. With Computer-Aided Diagnosis (CAD) system, time-consuming process in medical diagnosis\cite{RN106} can be accelerated. It can reduce the false negative rate as well. Hence, the development of an accurate and efficient computer system is crucial to medical imaging interpretation.

\par In most CAD systems, image segmentation is considered as a the pre-processing step\cite{4385141}. The quality and precision of segmentation can determine whether the computer analysis process will succeed to detect lung nodules. The key challenge of conventional lung segmentation methods is that it may fail to detect these nodules (candidate abnormalities)\cite{RN107}. While more complex algorithms can resolve this problem, they usually perform more computation. Typically, there is a trade-off between computing time and the quality of the segmentation.

\par Recently, graphic processing units (GPUs) have outperformed central processing units (CPUs) with their high parallelizability in many areas of high performance computing (HPC), including large-scale physics simulations, real-time weather forecasting, and all sorts of decision making in medicine, finance, manufacture, and beyond.  In order to improve the accuracy of segmentation, this study addresses the issue of intensive computation, proposes a framework to accelerate image segmentation, and reaches the speedup of the performance ratio of GPUs and CPUs.

\section{Image segmentation}

\par Image segmentation can distinguish the object of interest from its background\cite{RN105}. When the interested object has been detected, the segmentation should be stopped. The granularity of execution depends on the problem domain. For example, in the automatic verification of electronic packaging, the main objective is to determine whether there is any lack of components or damaged electrical wiring. There is no reason to do a finer identification that is smaller than these elements. For every computer vision based application such as autonomous driving, target tracking, tumor detection, texture extraction, and face recognition, requires a different scale of segmentation\cite{RN115}. Also, image segmentation usually demands a high quality and fast performance.

\par There are various methods to segment an image including edge detection, thresholding, region growing\cite{295913}, watershed\cite{641413}, active contours, level sets\cite{Zhang2010668} etc. Some of them are based on the contrast of nearby pixels, and others are based on the similarities of adjacent pixels. The following part will describe the basic ideas of edge detection, thresholding, and region growing.

\subsection{Edge detection}

\par Edge detection is a local image processing method designed to detect edge pixels. It is an approach used most frequently for segmenting images based on an abrupt change in intensities\cite{Gonzalez:2006:DIP:1076432}. There are three basic steps involved in edge detection.

\subsubsection{Smoothing}

\par
\begin{figure}[!t]
\includegraphics[width=\columnwidth]{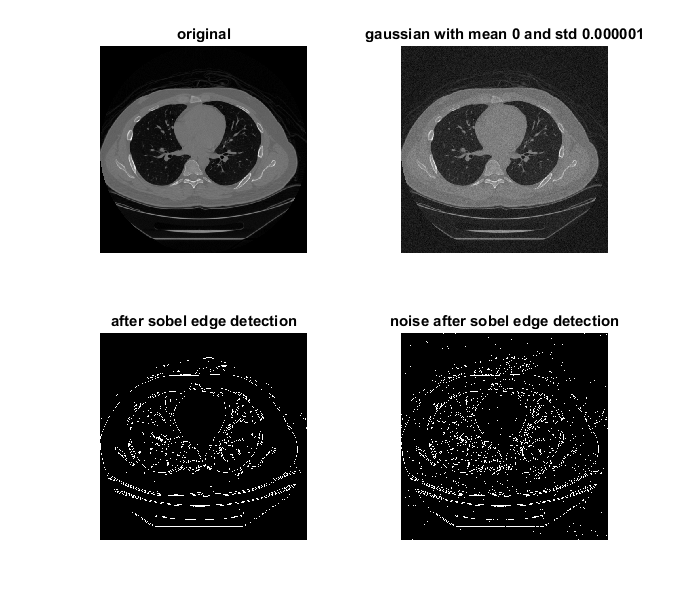}
\caption{Edge detection is sensitive to noise. Top left image is the original lung CT image, top right is the image after adding Gaussian white noise with mean $0$ and variance $0.000001$, bottom left is the result of edge detection with Sobel method\cite{aybar2006sobel} applied on the original image, and bottom right is the result of edge detection on the noisy image.}
\label{edge}
\end{figure}

 Edge detection usually involves the calculation of gradients (the partial difference of the image). Since random noise in an image can lead to shocks in the first derivative and the second derivative, noise can cause corruption to occur as shown in Figure \ref{edge}. Smoothing filters such as Gaussian smoothing filter are used to blur images and remove detail and noise.

\subsubsection{Detection of edges}

First or second-order derivative operator can extract all potential candidates to become edge points. By analyzing intensity profiles and different directions, edge points aggregate to an edge.

\subsubsection{Edge selection}

Edges are selected based on the strength and direction. Image gradient indicates the edge strength and the direction at location $(x, y)$ of an image $f$, which can be defined as a vector operator

$$\nabla f(x,y)\equiv grad(f(x,y))\equiv \begin{bmatrix}
       g_x           \\[0.3em]
       g_y \\[0.3em]
     \end{bmatrix} \equiv \begin{bmatrix}
    \frac{\partial f}{\partial x} \\[0.3em]
    \frac{\partial f}{\partial y} \\[0.3em]
     \end{bmatrix}
$$

\noindent where $f$ is the matrix of an image, $(x,y)$ an arbitrary location on the image, 
$g$ the gradient of intensity change in $x$ or $y$ direction. The vector $\nabla f$ gives an idea about the geometrical property and directs at the change rate of $f$ at location $(x, y)$\cite{Gonzalez:2006:DIP:1076432}.

Edges characterize boundaries between interested object and useless information. Thus, edge detection can segment images into foreground and background\cite{maini2009study}. However, noise will corrupt the result of such gradient-based approaches.

\subsection{Thresholding}

\par
\begin{figure}[!t]
\includegraphics[width=\columnwidth]{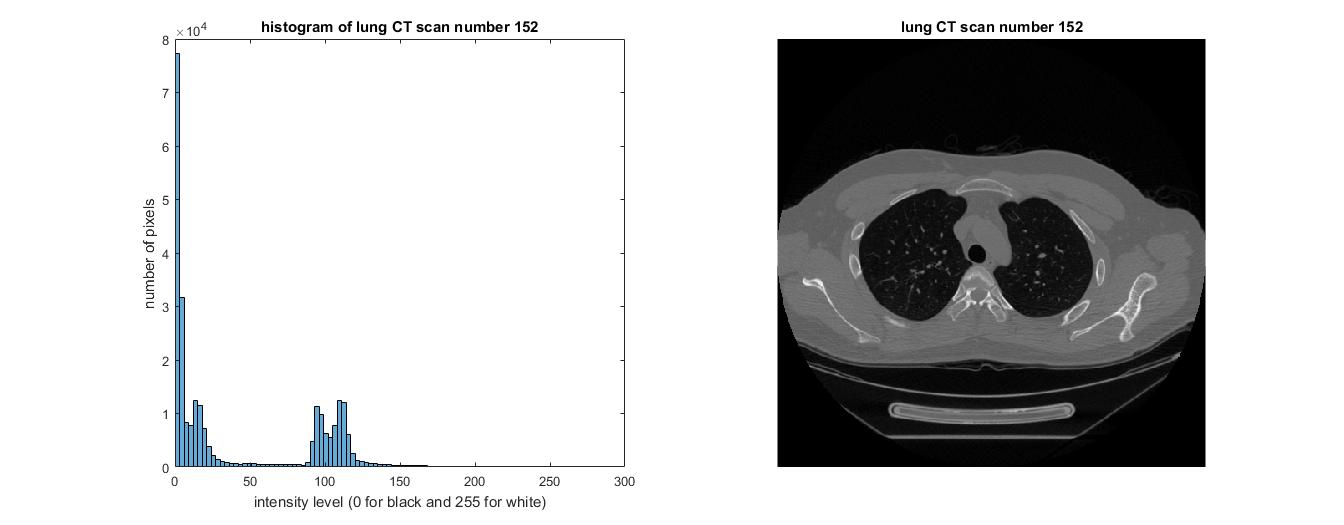}
\caption{Histogram of a lung CT image. Left hand is the histogram of the image at the right hand side. It is a 8-bit gray scale image after the pre-processing step. The x axis is the brightness of a pixel ($0-255$) and the y axis is the number of occurrence of each brightness.}
\label{hist}
\end{figure}

 Thresholding involves the step of selecting a threshold value in a measurement space (histogram) and labels every pixel into the foreground or background. Histogram of the brightness level is a common measurement space used in thresholding. One example of histogram of a lung CT scan is shown in Figure \ref{hist}, while most of the pixels in the CM scan are black (0 value on x axis). Thresholding uses brightness information and splits pixels into different classes. In the simplest form, a threshold value is determined to segment the image into interested objects and background with a thresholding kernel function in Algorithm \ref{kernel}.

\begin{algorithm}
\SetAlgoLined
\KwResult{Pixels in an image are classified as the foreground (1) or background(0)}
 Initialize image, threshold value T, x, y\;
 \ForEach{pixel (x,y) in the image}{
  \eIf{image(x, y) \textgreater= T}{
   image(x, y) = 1\;
   }{
   image(x, y) = 0\;
  }
 }
 \caption{Thresholding kernel function. $T$ is the threshold value. $(x,y)$ is the location of a pixel.}
 \label{kernel}
\end{algorithm}

\par Although, thresholding kernel function is relatively easy to be implemented. The key challenges are its inability to exclude salt and pepper noise and the selection of a proper threshold value.

\subsection{Region growing}

\par Region growing is an algorithm that groups pixels into regions. When adjacent pixels meet the predefined criteria, they group up as a region (foreground)\cite{Gonzalez:2006:DIP:1076432}. It starts with a set of seed points. Afterward, it grows regions by joining neighbouring pixels that meet these predefined foreground criteria.

\par Since the initialization may affect the subsequent growing regions, proper selection of seed points is important. A priori information about the image can help a lot in the selection of seed points. If no information is available, seed points can be initialized by clustering the brightness (intensities) of pixels.

\par Minimum area threshold and similarity threshold criteria depend not only on problems but also on types of the image. If the image is grayscale, region analysis must be performed with a set of descriptors based on the intensity levels and the spatial properties\cite{Gonzalez:2006:DIP:1076432}.

\par All of the image segmentation methods have their merits and drawbacks. This study addresses the problem of time-consuming segmentation in medical image processing. With a high data parallelism, low memory usage and GPU suitability\cite{RN10}, this study adopts thresholding. Two-dimensional histogram and maximum Tsallis entropy\cite{RN120} are used to determine the threshold value automatically. Comparing to the conventional thresholding methods, 2D histogram takes the spatial correlation of pixels into consideration and thus the quality of segmentation increases. However, computational tasks required to get the maximum Tsallis entropy\cite{RN121} are huge and the demand for computing power may grow fast with a higher resolution image. With the combination of thresholding, 2D histogram, and maximum Tsallis entropy, we propose a stepwise framework to render a fast image segmentation and accelerate compute-intensive tasks by using GPUs.

\section{Method of airway reconstruction}

\par
\begin{figure}[!t]
\includegraphics[width=\columnwidth]{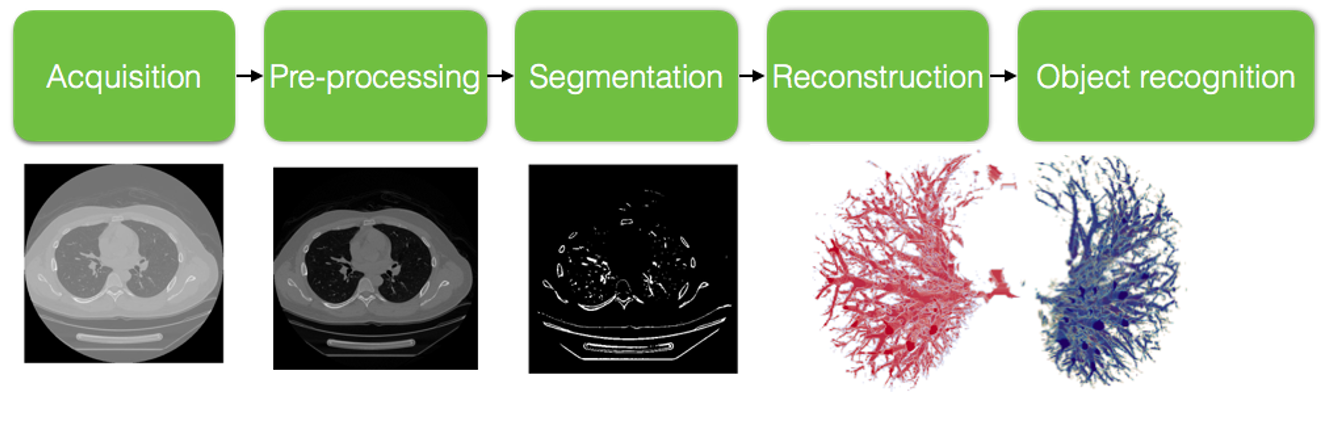}
\caption{Five steps of airway reconstruction. 3D airway reconstruction involves five steps. Acquisition retrieves of CT scan from CT/MRI machine. Pre-processing rescales the Hounsfield unit (HU) of CT scans to a typical 8-bit image. Segmentation classifies pixels into airway part and non-airway part. Reconstruction models the 3D airway from previous steps. Object recognition usually involves human diagnosis.}
\label{steps}
\end{figure}

 An airway reconstruction involves five steps, including the image acquisition, pre-processing, image segmentation, reconstruction, and object recognition, as shown in Figure \ref{steps}.

\subsection{Image acquisition}

\par
\begin{figure}[!t]
\includegraphics[width=\columnwidth]{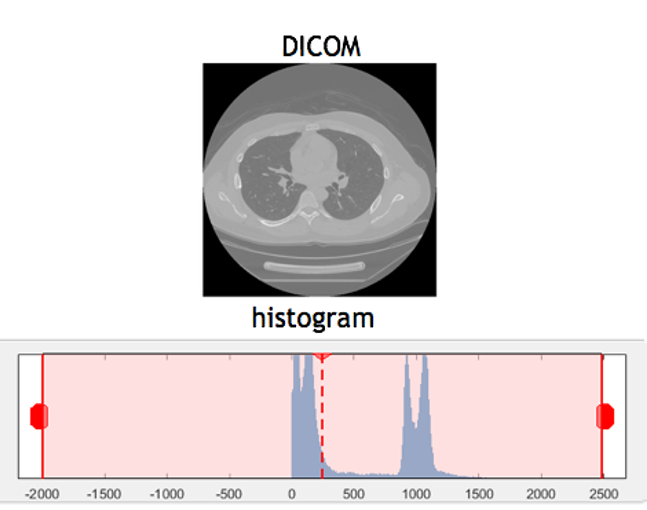}
\caption{Input DICOM. Top is the direct visualization of a DICOM. Bottom is the histogram of this image. The two red limits denote the maximum and minimum values of the Hounsfield Unit (HU) in this scan. Background is the black region outside of X-ray detector with HU value -2000.}
\label{original}
\end{figure}

 Image acquisition is a process of retrieving Computed Tomography (CT) scans of a patient. Digital Image and COmmunications in Medicine (DICOM) is the standard format for handling, storing, printing, and transmitting information about medical imaging. The histogram of a DICOM image ranges from -2000 to 2500 as shown in Figure \ref{original}, while the number stands for the Hounsfield scale (CT number) which is used to describe radiant intensity. A Hounsfield Unit ($HU$) scale given below denotes the linear transformation of the original attenuation coefficient measurement into the one in which the radiant intensity of distilled water at standard pressure and temperature,
\begin{gather*}
    HU=1000\times\frac{\mu-\mu_{water}}{\mu_{water}-\mu_{air}}
\end{gather*}
\noindent where $HU$ is the Hounsfield unit, $\mu$ the radiant intensity of a tissue, $\mu_{air}$ the linear attenuation coefficients of air and $\mu_{water}$ the linear attenuation coefficients of water.

\subsection{Pre-processing}

\par
\begin{figure}[bhp]
\includegraphics[width=\columnwidth]{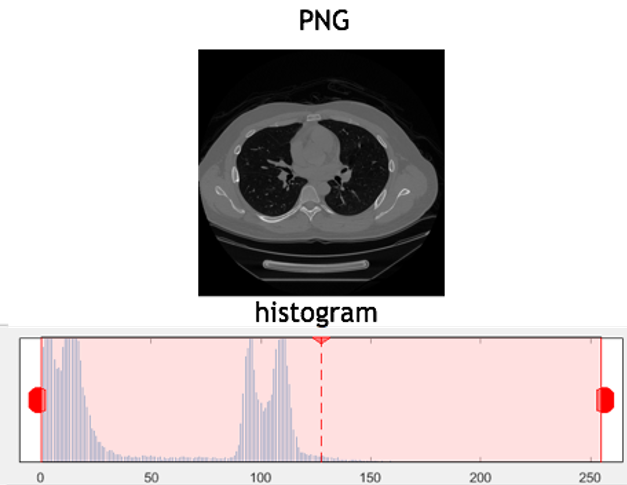}
\caption{Image after pre-processing. Top is the image after rescaling from the Hounsfield Unit (HU) to a typical 8-bit grays scale image. Bottom is the corresponding histogram.}
\label{pre}
\end{figure}

 In a typical gray scale image, there exist only brightness levels ranging from 0 to 255. Here a pre-processing step is applied to change the value of -2000 (outside of the X-ray detectors) to 0 (brightness level of the background). Then all intensity levels are linearly transformed to the range 0 to 255, as shown in Figure \ref{pre}.

\subsection{Segmentation}

\par This study combines algorithms to automatically segment the airway part in an image, including morphology operations, thresholding, and threshold value selection by Tsallis entropy. Segmentation is the most computationally intensive part in the building block of 3D airway reconstruction.

\subsubsection{Morphology}

\par Morphology opening can create a mask that filters out chest tissues and bones except for airway. Opening is derived from the fundamental operations: erosion and dilation. It is noted that the opening is the dilation after the erosion of the set $A$ by a structure element $B$. Here $A$ is the gray scale image array and $B$ is the shape used to interact with a given image $A$. Structure elements are usually small and simple. Here, erosion is defined as the notation

\begin{gather*}
    A\ominus B=\{z|(B)_z \subseteq A\}
\end{gather*}
\noindent where $A$ is a matrix of pixels, $B$ a structure element, $\ominus$ the erosion operator, and $z$ the translation of $B$, which is contained in $A$. By definition, the erosion of A by B is the set of all points z such that B, translated by z, is contained in A. And dilation is defined as the notation 

\let\oldemptyset\emptyset
\let\emptyset\varnothing

\begin{gather*}
    A\oplus B=\{z|(\hat{B})_z \cap A \neq \emptyset\}\\
\end{gather*}
\noindent where $A$ is a matrix of pixels, $\hat{B}$ a symmetric structure element of $B$, $\oplus$ the dilation operator, $z$ the displacement of $\hat{B}$ overlapping with $A$. Note that $A\oplus B$ is the set of all displacements, z, such that \(\hat{B}\) and A are overlapped by at least one element \cite{Gonzalez:2006:DIP:1076432}. While erosion will shrink objects in an image, dilation can grow and thicken the object in an image. As a consequence, opening removes the small noise by erosion first and grows objects to their original size. The definition of opening is

\begin{gather*}
    A\bigcirc B=(A\ominus B)\oplus B
\end{gather*}
\noindent where $A$ is an image, $B$ a structure element, $\bigcirc$ opening, $\ominus$ erosion, $\oplus$ dilation.

\par
\begin{figure}[bhp]
\includegraphics[width=\columnwidth]{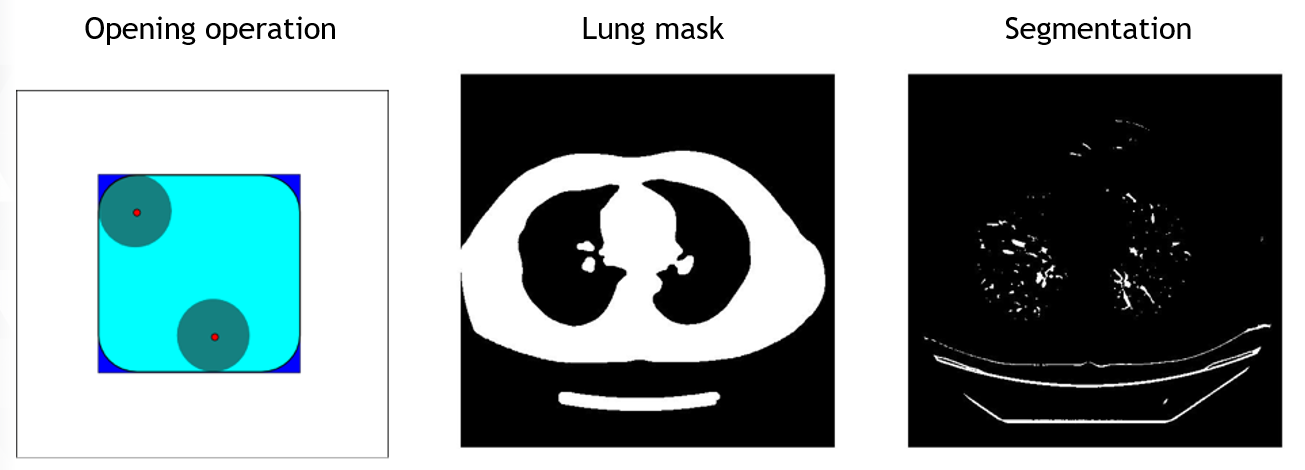}
\caption{Morphology operation. Left is the illustration of an opening operation. Middle is the chest mask created by applying the opening operation on the pre-processed image. Right is the image after applying the chest mask on the pre-processed image.}
\label{mor}
\end{figure}

 With the help of opening operation and a disk-like structure element of size 10 pixels, the chest can be removed by subtracting the original image from the opened mask as shown in Figure \ref{mor}.

\subsubsection{Thresholding}

After morphology opening operation, white bones, gray chest tissues, and other noises still remain. Thresholding is applied and the threshold value is automatically selected by maximizing the Tsallis entropy \cite{RN120}.

\subsubsection{Tsallis entropy}

\par Tsallis entropy uses Shannon entropy\cite{Shannon:2001:MTC:584091.584093} originated from the information theory that considers the gray level image as a probability distribution\cite{RN121}. It has the advantage of using global objective properties and its ease of implementation. Additionally, Tsallis $\alpha$ coefficient can be adjusted and plays a critical role in tuning parameters for better performance. Thresholding has been criticized for the ignorance of the spatial relationship between pixels since conventional thresholding involves only splitting the level of brightness. Here, this study uses two-dimensional Tsallis entropy\cite{RN120} which takes the spatial information into consideration by calculating the two-dimensional histogram. The first dimension of the histogram has a horizontal axis representing the levels of brightness and a vertical axis representing the number pixels in that particular level. The second dimension is the histogram derived from the average images addressing the effect of neighbor pixels. The average gray value for the $3*3$ neighborhood of each pixel is calculated as

\begin{gather*}
    g(x, y)=\lfloor \frac{1}{9}\sum_{i=-1}^{1}\sum_{j=-1}^{1}f(x+i, y+i)\rfloor
\end{gather*}
\noindent where $g$ is the average gray image, and $(x,y)$ is any location on the image.

Note that the symbol \(\lfloor \quad \rfloor\) denotes the decimal part of the number. A 2D histogram can be constructed using the probability of the original image and the average image.  A 2D histogram can be defined as

\begin{gather*}
    h(i, j)=prob(f(x, y) = i \& g(x, y) = j)\\
\end{gather*}
where $h$ is the 2D histogram of an image and its average image, $(x,y)$ an any location on the image, and $i,j$ the brightness level.

The 2D histogram plane is divided into four regions by a threshold value pair $(t, s)$, where $t$ is a threshold for pixel intensity of the original image and $s$ is another threshold for its average image. 1\textsuperscript{st} and 3\textsuperscript{rd} quadrants represent respectively the objects and the background. 2\textsuperscript{nd} quadrant can be ignored as these two contain information mainly about edges and noises.
\par Let $P$ be the set of the possibilities of brightness levels $={p_0,p_2,p_3,...,p_{n - 1}}$ where $p_0, p_2, ..., p_{n - 1}\geq 0$ and $\sum_{s=0}^{n - 1}{p_s}=1$. The entropy of splitting the image into background (class $1$) and foreground (class $2$) by this threshold value pair $(t, s)$ can be defined as 

\begin{gather*}
    H^{\alpha}_1=\frac{1}{\alpha-1}[1-\sum_{i=0}^{t}{\sum_{j=0}^{s}{(\frac{P(i)}{P_1(t)})^\alpha}}]\\
    H^{\alpha}_2=\frac{1}{\alpha-1}[1-\sum_{i=t+1}^{L-1}{\sum_{j=s+1}^{L-1}{(\frac{P(i)}{P_2(t)})^\alpha}}]
\end{gather*}
\noindent where $H^\alpha_1$ is the Tsallis entropy of background (class $1$), $\alpha$ a tunable parameter, $t$ the threshold on 1D histogram, $s$ the threshold on 2D histogram, $H^\alpha_2$ the Tsallis entropy of foreground (class $2$), and $L$ the length of the level variation. In the above equations,

\begin{gather*}
    P_1(t, s)=\sum_{i=0}^{t}{\sum_{j=0}^{s}{ p_{i,j}}}\\
    P_2(t, s)=\sum_{i=t+1}^{L-1}{\sum_{j=s+1}^{L-1}{ p_{i,j}}}
\end{gather*}
\noindent where $p_{i,j}$ is the possibility of the brightness $i$ in the original image and brightness $j$ in its average image, $t$ the threshold on first dimension, $s$ the threshold on second dimension, and $L$ the length of the level variation. The best threshold value pair $(t, s)$ for 2D Tsallis entropy can be obtained by maximizing the value of \(\varphi_\alpha\) given below 

\begin{gather*}
    \varphi_\alpha(t, s)=Arg\quad max([H_1^\alpha(t, s)+H_2^\alpha(t, s)\\
    +(1-\alpha)H_1^\alpha(t, s)H_2^\alpha(t, s)]),\\
\end{gather*}
\noindent where $H_1$ is the Tsaiils  entropy of class $1$, $H_2$ the Tsaiils entropy of class $2$, $\alpha$ a tunable parameter, $t$ the threshold on 1D histogram, and $s$ the threshold on 2D histogram. Finally, only $t$ is used as the theshold value.

\subsection{3-D reconstruction}

\par
\begin{figure}[!t]
\includegraphics[width=\columnwidth]{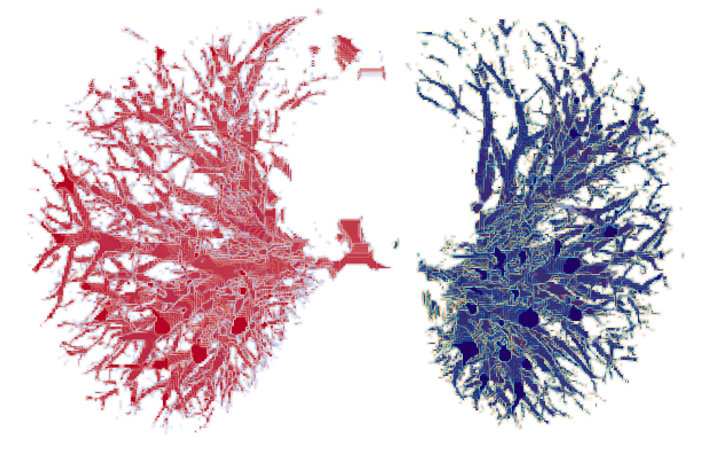}
\caption{Airway after reconstruction from the top view. Red part is the airway in the left lung. Blue part is the airway in the right lung.}
\label{rec}
\end{figure}

 After segmentation, there is a series of CT images containing only the airway structure. When stacking all the scans together with appropriate slice intervals defined in DICOM, 3D model of an airway can be constructed. Figure \ref{rec} shows the reconstructed airway from the top view. 

\subsection{Object recognition}

\par Finally, the 3D model can be used for other object recognition tasks, such as doctor diagnosis or serving as the input for a nodules analysis.

\begin{figure}[!t]
\includegraphics[width=\columnwidth]{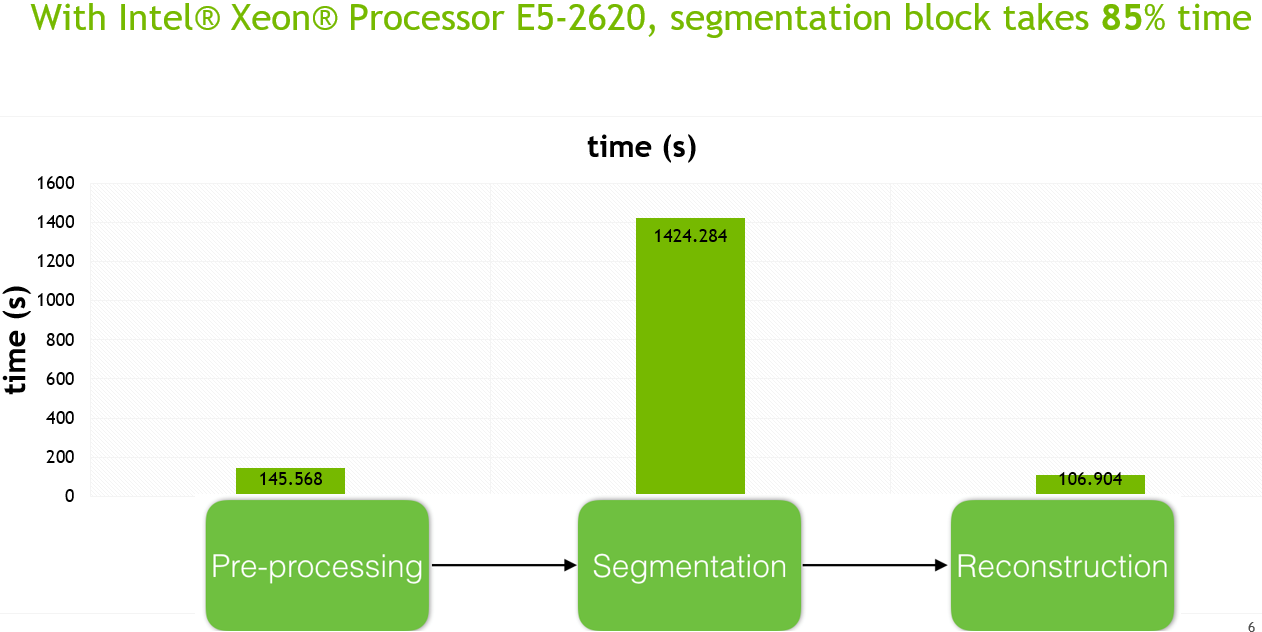}
\caption{Run time for each step in 3D airway reconstruction. Segmentation takes up 85\% of the time.}
\label{time}
\end{figure}

\par Among all steps of 3D airway reconstruction, segmentation is the most time consuming part among all the building blocks shown in Figure \ref{time}. Image segmentation has taken nearly 85\% of running time (20 minutes out of 26 minutes). The calculation on 2D Tsallis entropy is responsible for this. In the following paragraph, this study will explain why GPU has a great potential for accelerating this algorithms.

\section{CPU/GPUs architecture}

\par With the growing demand on processing power, the traditional CPU architecture can no longer catch up with the data-generating speed\cite{RN116}. Computing and analyzing the exploding data in real time become important aspects in many scientific computing domains. For example, a finer grid of weather stations can collect more data, including temperature, humid, and pressure etc, while these may be useless if the forecasting model can't finish computation fast enough. Fortunately, with the advance in Graphic Processing Units (GPUs), which were originally designed to process animation or image on screen, now General Purpose Graphic Processing Units (GPGPUs) can be used in practical computation. In the past few years, there have been many studies reporting that GPUs can gain a considerable speedup\cite{RN8}\cite{RN10}\cite{RN116}. To understand why CPU and GPU can make a big difference in performance, this study dives into the design of them. 

\par Fundamentally, CPU and GPU are built for different purposes. CPU is designed for a wide range of applications. It is good at giving a quick response to every single task. CPU techniques such as branch prediction, out-of-order execution, and super-scalar\cite{RN116} have been developed. These developments come, however, with a huge cost of power consumption and high complexity. As a consequence, today CPUs can only be allowed to have a small number of cores to prevent thermal problem and reduce energy consumption. On the other hand, GPUs are used in rendering a large amount of data for parallelization since pixels on the screen can be computed independently. A number of cores (thread processors) share streaming multiprocessors which work as the control unit to dispatch instructions to these cores\cite{RN10}. As a result, GPUs have intrinsic parallelization ability in the hardware architecture when comparing to CPU. GPUs, on the other hand, don't address issues on the cache size, memory size and performance of thread switching (context switching).

\section{Result and discussion}

\par Code runs on CPU, while chest mask filtering, threshold value selection by maximizing Tsallis entropy, and thresholding are programmed in a sequential manner.


\par From the hardware environment comparison in Table \ref{table:hard}, it can be seen that GPUs have more 4 to 5 hundred times of cores than CPUs. Moreover, the peak single precision floating point operation per second is 30 to 40 times of the CPU one.
    
    \begin{table}
    \caption{Hard environment}
    \label{table:hard}
    \begin{center}
    \begin{tabular}{ | l | l | l |}
    \hline
    Card & Cores & Peak single precision flops \\ \hline
    Nvidia Tesla K40c & 2880 & 4.29Tflops \\ \hline
    Nvidia Tesla K20c & 2496 & 3.52 Tflops \\ \hline
    Intel Xeon E5-2630 & 6 & 0.134 Tflops \\
    \hline
    \end{tabular}
    \end{center}
    \end{table}
    
\subsection{Preliminary implementation}
\par In the first experiment, a novel approach that contains only parallelized for loop in the calculation of Tsallsi entropy. Performance assessment is made on comparing the sequential and parallellized one on CPU and the parallellized code on GPU. Simulated results are tabulated in Table \ref{table:pre}. After the implementation of parallelism concept, the parallellized code enabled a speedup gain of 6 on CPU and 9 on GPU. Here, this study can make the first conclusion that parallelism is a must in the era of multi-core computers available everywhere and an easy parallel algorithm can offer a considerable performance improvement.

\par Since the specification and benchmark have shown a larger difference in computing power in terms of the peak single precision FLOPS (float point operation per second), this study has gone further and conducted the next experiment that involves the tuning of the memory setting and changing the grid and block size.
    
    \begin{table}
    \caption{Assessment study for the preliminary implementation of parallelization on a CPU and a GPU}
    \label{table:pre}
    \begin{center}
    \begin{tabular}{ | l | l | l | l | }
    \hline
    Processor & Language & Time(sec) & Note \\ \hline
    Intel Xeon & NA & 158.08 & Sequential code \\ \hline
    Intel Xeon & OpenMP & 6.36 & Double checked locking \\ \hline
    Nvidia Tesla K20c & CUDA & 9.34 & Global memory only\\
    \hline
    \end{tabular}
    \end{center}
    \end{table}
    
\subsection{Fine tuning on memory arrangement}
    \par There are many types of memory in GPU. Each of them has its own advantages and drawbacks. For global memory having a slower access speed, it has, however, a larger space. For shared memory possessing a faster speed, there is only a limited size available to store arrays and variables. Here, this study compares different approaches by using different kinds of memory in terms of global, shared, texture memory. The result is shown in Table \ref{table:memory}. It can be seen that texture memory has the best performance. This assessment study leads to our second conclusion. Despite of faster performance, texture memory renders a lower accuracy. In computational science, accuracy is of great importance. Shared memory is therefore more preferable.
    
    \begin{table}
    \caption{Performance comparison for different memory arrangements}
    \label{table:memory}
    \begin{center}
    \begin{tabular}{ | l | l | l |}
    \hline
    Memory usage on Tesla K20c & Run time(sec) & Speedup gain \\ \hline
    Global memory & 3 & 1 \\ \hline
    Shared and global memory & 0.471 & 6.36 \\ \hline
    Texture and global memory & 0.321 & 9.34 \\
    \hline
    \end{tabular}
    \end{center}
    \end{table}
    
\subsection{Fine tuning on the grid and block setting}
    
    \par GPUs have another interesting configuration about the grid and block settings. Associated with the thread scheduler, different grid and block sizes can actually have impact on the speedup gain. From the result in Table \ref{table:blockgrid}, it can be seen that if a GPU don't create enough threads then the performance would be deteriorated by a factor of ten. On the other hand, if a GPU keeps increasing the number of threads the speedup gain wouldn't be improved any more. Moreover, simulation would suffer from communication overhead. So far, this study may draw the third conclusion. Tuning block and thread numbers aids optimizing the performance. Let each thread do less job.
    
    \begin{table}
    \caption{Different performances with different block and grid settings}
    \label{table:blockgrid}
    \begin{center}
    \begin{tabular}{ | l | l | l |}
    \hline
    Grid size / block size & run time(ms) on Tesla K40c & Tasks per thread \\ \hline
    32 / 32  & 3714.5 & $2^6$ \\ \hline
    128 / 128 & 335.5 & $2^2$ \\ \hline
    256 / 256 & 388.1 & $2^0$ \\ \hline
    512 / 512 & 474.9 & \textless$2^0$\\
    \hline
    \end{tabular}
    \end{center}
    \end{table}

\subsection{Implementation on multiple GPUs (2*K40)}

\par In order to explore the limitation on the acceleration of GPUs, this study also test the performance of multiple GPUs. With the combination of strategies to select the best memory and grid and block settings, GPU will have to modify the job distribution from a single hardware to communication between multiple GPUs and the reduction of results from different devices. While communication and merging of the final result would require an additional work and message exchanging, the overhead cannot be neglected. Table \ref{table:multiple} shows the final result of the integrated result from the previous discussion. It can be observed that there is an additional gain of 1.5 times speedup in the two-GPU version.
    
    \begin{table}
    \caption{Performance of the combination of the best memory, block and grid settings with multiple GPUs}
    \label{table:multiple}
    \begin{center}
    \begin{tabular}{ | l | l | l |}
    \hline
    Platform & Run time(sec) & Speedup \\ \hline
    CPU & 14.335 & 1 \\ \hline
    CPU + 1 GPU & 0.335 & 43 \\ \hline
    CPU + 2 GPU & 0.232 & 61.8 \\
    \hline
    \end{tabular}
    \end{center}
    \end{table}

\par To sum up, this study have attempted different arrangements of the memory type and the grid and block size. The best performance can be obtained by the following two principles.

\begin{enumerate}
  \item Despite of faster performance, texture memory renders a lower accuracy. While in computational science, accuracy is of great importance, shared memory is more preferable.
  \item Tuning the block and thread numbers aids optimizing the performance. Let each thread do less job.
\end{enumerate}

\par As this study starts to involve more GPUs in the calculation, the overhead of communication cannot be ignored. This study can draw a conclusion from these experimental results as follows. While estimating the speedup in parallelized algorithm, the best performance can reach the theoretical value (in terms of the peak single floating operation per second), this study can have 40 times speedup in this case study.

\section{Concluding remarks}

\par Computational algorithms usually have a trade off between complexity and performance. In other words, an algorithm which has high accuracy and high precision needs more computing time. However, there is always a need to improve algorithms in terms of the accuracy and precision. In order to cope with the problem of the increasing computational time, it would need a new paradigm for designing an effective algorithm.
\par There goes a saying that you can't make brick without straw. As this study develops a better algorithm, the platform (hardware) also plays a crucial role in a real application. Recently, GPU has achieved a great performance improvement with regard to FLOPS (float point operation per second) via a large amount of cores and SIMD (single instruction and multiple data). 
\par This study has proposed a novel framework to get the speedup concerning the performance ratio of CPUs and GPUs, assess different arrangements of the grid and block and memory settings and recommend the best configuration for a specific algorithm. Specifically, in our acceleration of image segmentation using Tsallis entropy, this study has performed a series of experiments on hyperparameter of CUDA runtime API (application interface). The conclusions of this study are drawn as follows. Firstly, tuning the block and thread numbers aids optimizing the image segmentation performance. The guide one should follow is that let each thread do less job. Secondly, despite of gaining a faster performance, texture memory renders a lower accuracy. While in computational science, accuracy is of great importance, shared memory is more preferable. Thirdly, more GPUs help to accelerate computation while suffering overheads from job distribution and communication.

%
\IEEEpeerreviewmaketitle

\section*{Acknowledgment}

The authors would like to thanks Neo Shih-Chao Kao for providing his own chest CT images. National Taiwan University Computer Center provides high performance computing environment with Nvidia K80 and K40 GPU cards.

\ifCLASSOPTIONcaptionsoff
  \newpage
\fi



%

\bibliographystyle{ieeetr}

%





\end{document}